\documentclass[%
preprint,
 amsmath,amssymb,
 aps,
]{revtex4-2}

\usepackage{graphicx}% Include figure files
\usepackage{dcolumn}% Align table columns on decimal point
\usepackage{bm}% bold math
\begin{document}

%\preprint{APS/123-QED}

\title{Method of filtration in first passage time problems}% Force line breaks with \\

\author{Yuta Sakamoto}
 \altaffiliation{Department of Physical Sciences, Aoyama Gakuin University, 5-10-1 Fuchinobe Chuo-ku,Sagamihara, Japan}%Lines break automatically or can be forced with \\
 \email{sakamoto2021@phys.aoyama.ac.jp}
\author{Takahiro Sakue}%
 \altaffiliation{Department of Physical Sciences, Aoyama Gakuin University, 5-10-1 Fuchinobe Chuo-ku,Sagamihara, Japan}%Lines break automatically or can be forced with \\
 \email{sakaue@phys.aoyama.ac.jp}
  
\date{\today}% It is always \today, today,

\begin{abstract}

Statistics of stochastic processes are crucially influenced by the boundary conditions. 
In one spatial dimension, for example, the first passage time distribution in semi-infinite space (one absorbing boundary) is markedly different from that in a finite interval with two absorbing boundaries. 
Here, we propose a method, which we refer to as a method of filtration, that allows us to construct the latter from only the knowledge of the former. 
We demonstrate that our method yields two solution forms, a method of eigenfunction expansion-like form and a method of image-like form.
In particular, we argue that the latter solution form is a generalization of the method of image applicable to a stochastic process for which the method of image generally does not work, e.g., the Ornstein-Uhlenbeck process.

\end{abstract}
\maketitle

\section{Introduction}
Statistics of first passage time (FPT) underlie various phenomena encountered in natural sciences, technological applications, and our daily lives [1-11].
Being a stochastic quantity, the FPT is characterized by its distribution, and its computation constitutes an imperative subject in the theory of stochastic processes.
Typically, one asks how long it takes for random walkers, starting from a specified initial position, to reach the target location, represented as an absorbing boundary.
Therefore, a problem definition requires the boundary condition, which significantly impacts the solution.

Suppose, for example, you have bought stock in a company at a price $x_0$ today ($t=0$). 
The stock price is subject to daily fluctuations in the market, so you may be interested in how many days it takes to achieve the target price $x_h (> x_0)$.
The relevant quantity here is the FPT addressing when the price hits the target for the first time.
However, you may employ a strategy with a lower limit of $x_{\ell}$ for the allowable price; once the price reaches $x_{\ell} ( < x_0)$, you will sell the stock, and the game is over. 
In this case, the quantity of interest is how long it takes for the price to hit the target $x_h$ without hitting the tolerance limit $x_{\ell}$.

In terms of theoretical formulation, while the first case deals with a semi-infinite system with one absorbing boundary, the second case limits the system to a finite interval by two absorbing boundaries.
The FPT distributions of these two cases would generally differ on a qualitative level, and little is known about their relations.
This difference is ascribable to the fact that the standard methods of solution vary with the type of boundary conditions \cite{guide}.
In this paper, we address this issue for homogeneous Markovian processes.
We propose a new method of solution for the latter problem with two absorbing boundaries, in which the solution is constructed only from the solutions for the former problem with one absorbing boundary.

\section{Preliminaries}
In this section, we introduce the stochastic model we deal with in this paper and a standard method of solution in the system with one absorbing boundary as preparation for the new method we suggest. 

%%%%%%%%%%%%%%%%%%%%%%%%%%%%%%%%%%%%%%%%%%%%%%%%%%%%%%%%%%%%%%%%%%%%%%%%%%%%%%%%%%%%%%%%%%%%%%%
%%%%%%%%%%%%%%%%%%%%%%%%%%%%%%%%%%%%%%%%%%%%%%%%%%%%%%%%%%%%%%%%%%%%%%%%%%%%%%%%%%%%%%%%%%%%%%%
% 2-1
\subsection{Stochastic model}
We consider one-dimensional Markovian dynamics described by the overdamped Langevin equation
\begin{equation}
\label{21_Langevin}
  \frac{dx}{dt} = -\frac{1}{\gamma} \frac{dU(x)}{dx} + \xi(t) ,
\end{equation}
where $\gamma$ is the friction constant, $U(x)$ is the potential energy and $\xi(t)$ is white Gaussian noise with zero mean and autocorrelation
\begin{equation}
\label{21_correaltion}
\langle \xi(t) \xi(s) \rangle = 2 D \delta(t-s).
\end{equation}
Here, $D$ is the diffusion coefficient, and $\delta(t)$ represents the delta function. 
In this paper, we deal with the Brownian motion model in flat potential $U(x)=0$ (free diffusion), linear potential $U(x)=\alpha x$ (biased diffusion), and harmonic potential $U(x)=\frac{k}{2} (x-a)^2$ (Ornstein-Uhlenbeck process).

Let $P(x,t|x_0,t_0)$ be transition probability, which is the probability density for a random walker positioned $x_0$ at time $t_0$ to transit $x$ at time $t$.
Its time evolution is described by the Fokker-Plank equation \cite{FPequ}
\begin{eqnarray}
\label{21_FP}
  \frac{\partial P(x,t|x_0,t_0)}{\partial t} = L_{FP}(x) P(x,t|x_0,t_0)
\end{eqnarray}
with the operator
\begin{eqnarray}
\label{21_LFP}
  L_{FP}(x) = \frac{\partial}{\partial x} \left[ \frac{1}{\gamma} \frac{\partial U(x)}{\partial x} \right]
              + D \frac{\partial^2}{\partial x^2}.
\end{eqnarray}
%

%%%%%%%%%%%%%%%%%%%%%%%%%%%%%%%%%%%%%%%%%%%%%%%%%%%%%%%%%%%%%%%%%%%%%%%%%%%%%%%%%%%%%%%%%%%%%%%
%%%%%%%%%%%%%%%%%%%%%%%%%%%%%%%%%%%%%%%%%%%%%%%%%%%%%%%%%%%%%%%%%%%%%%%%%%%%%%%%%%%%%%%%%%%%%%%
% 2-2
\subsection{FPT in one boundary}

Let us set an absorbing boundary at $x=B$, and consider the probability density $F_{I}(t;x_0 \Rightarrow B)$ of FPT, where the right arrow with double lines indicates the first passage process from the initial position $x_0$ to the absorbing boundary $x=B$, and the subscript ``$I$" stands for the presence of {\it one} absorbing boundary that is imposed in the process under consideration.
For Markovian processes, it is known that the following relation \cite{guide} holds 
\begin{equation}
\label{22_conv1}
  P(B,t|x_0,0) = \int_{0}^{t} d\tau F_{I}(\tau;x_0 \Rightarrow B) P(B,t|B,\tau),
\end{equation}
where $P(x,t|x_0,0)$ is the solution of Eq.(\ref{21_FP}) with initial condition $P(x,0|x_0,0)=\delta(x-x_0)$ in infinite space, i.e. without boundary.
Noting that, for homogeneous processes, the right-hand side takes the convolution form, we obtain the solution for $F_I(t;x_0 \Rightarrow B)$ in the Laplace domain as
\begin{equation}
\label{22_conv_Lap}
 \tilde{F}_I(s;x_0 \Rightarrow B) = \frac{\tilde{P}(B,s|x_0,0)}{\tilde{P}(B,s|B,0)},
\end{equation}
where ${\tilde f}(s) = {\mathcal L}[f(t)] \equiv \int_0^{\infty} dt f(t) e^{-st}$ is the Laplace transform of $f(t)$. 
It is often useful to write 
\begin{equation}
\label{22_P_Lap}
 \tilde{P}(B,s|x_0,0) = C(B,s) u_s(x_0),
\end{equation}
where $u_s(x_0)$ is a solution of the homogeneous equation
\begin{equation}
\label{22_Eq_v}
 \left( L^{+}_{FP}(x_0) - s \right) u_s(x_0) = 0
\end{equation}
with $L^{+}_{FP}$ being the adjoint operator of $L_{FP}$ and the coefficient $C(B,s)$ is determined by boundary conditions \cite{fpt_on}.
By performing inverse Laplace transform ${\mathcal L}^{-1}$, 
we obtain
\begin{equation}
\label{22_conv_ILap}
F_I(t;x_0\Rightarrow B) = \mathcal{L}^{-1} 
                         \left[ \frac{\tilde{P}(B,s|x_0,0)}{\tilde{P}(B,s|B,0)} \right] 
                       = \mathcal{L}^{-1} 
                         \left[ \frac{u_s(x_0)}{u_s(B)} \right] .
\end{equation}
%

% 3-1
\section{\label{sec:level2-1}Method of filtration}
Now we set $B=0$, and an additional absorbing boundary at $x=L$, and consider the first passage process in the interval $x \in [0,L]$. 
A quantity of interest is the FPT, defined as the time for a random walker initially positioned at $x_0$  to reach the absorbing boundary at $x=0$ for the first time without touching the opposite boundary at $x=L$.
Let $F_{I\hspace{-2.5pt}I}(t;x_0 \Rightarrow 0)$ denote the probability density of the FPT, 
where the subscript ``$I\hspace{-2.5pt}I$'' refers to the presence of {\it two} absorbing boundaries. 
Our aim is to construct the solution of $F_{I\hspace{-2.5pt}I}(t;x_0 \Rightarrow 0)$ in terms of $F_I(t;x_0 \Rightarrow 0)$.

\begin{figure}[t]
\begin{center}
\includegraphics[width=150mm]{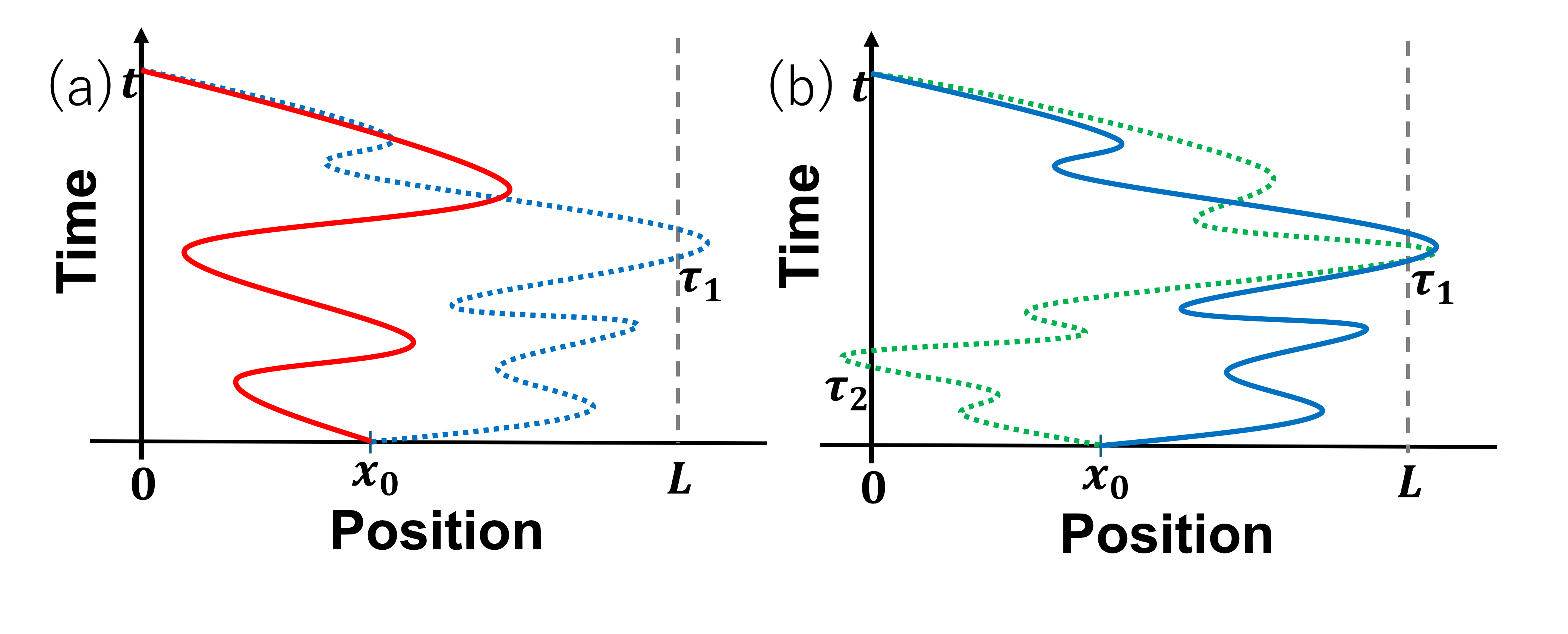}
\caption{Sketch of trajectories for a random walker to reach the absorbing boundary $0$.
(a) solid line: A trajectory that a random walker reaches the absorbing boundary $0$ at time $t$ for the first time without touching the opposite boundary $L$.
dashed line: A trajectory that a random walker touches the absorbing boundary $L$ for the first time at time $\tau_1$ before reaching the opposite boundary $0$ at time $t$.
(b) solid line: the same trajectory as the dashed line of Fig.(a). 
dashed line: A trajectory that a random walker touches boundary 0 at first at time $\tau_2$, then boundary $L$ at time $\tau_1(>\tau_2)$, and finally reaches boundary 0 at time $t(>\tau_1)$.}
\label{fig_trajectory}
\end{center}
\end{figure}

First, let us consider the relation of two FPT problems with different boundary conditions.
The first step is to focus on the question of why the solution with one boundary cannot be the solution with two boundaries.
Looking at trajectories until a random walker reaches absorbing boundary 0 for the first time, we realize that some of them touch opposite boundary $L$ before that (Fig.\ref{fig_trajectory}(a)).
In other words, the solution with one boundary includes the {\it wrong} trajectories unsuitable for the solution with two boundaries, which causes overestimation.
Therefore, by {\it filtering} such trajectories, we can pick out only the {\it right} trajectories suitable for the solution with two boundaries.
We can thus express the solution with two boundary $F_{I\hspace{-2.5pt}I}(t;x_0 \Rightarrow 0)$ as follows:
\begin{eqnarray}
\label{31_Mf1_x0to0}
F_{I\hspace{-2.5pt}I}(t;x_0\Rightarrow0) =
  F_I(t;x_0\Rightarrow0) - 
  \int_{0}^{t} d\tau F_{I\hspace{-2.5pt}I}(\tau;x_0\Rightarrow L) F_I(t-\tau;L\Rightarrow0), 
\end{eqnarray}
where the second term on the right hand in Eq.(\ref{31_Mf1_x0to0}) is the FPT probability that a random walker touches boundary $L$ at first but without touching another boundary 0, then reaches 0 at time $t$ for the first time.
Here, we still have a solution with two boundaries $F_{I\hspace{-2.5pt}I}(t;x_0 \Rightarrow L)$, which can be, however, expressed as, again using the filtration as follows:
\begin{eqnarray}
\label{31_Mf1_x0toL}
F_{I\hspace{-2.5pt}I}(t;x_0\Rightarrow L) = 
  F_I(t;x_0\Rightarrow L) - \int_{0}^{t} d\tau F_{I\hspace{-2.5pt}I}(\tau;x_0\Rightarrow 0) F_I(t-\tau;0\Rightarrow L) .
\end{eqnarray}
%%%%%%%%%%%%%%%%%%%%%%%%%%%%%%%%%%%%%%%%%%%%%%%%%%%%%%%%%%%%%%%%%%%%%%%%%%%%%%%%%%%%%%%%%%%%%%%%
%%%%%%%%%%%%%%%%%%%%%%%%%%%%%%%%%%%%%%%%%%%%%%%%%%%%%%%%%%%%%%%%%%%%%%%%%%%%%%%%%%%%%%%%%%%%%%%%
Substituting Eq.(\ref{31_Mf1_x0toL}) into Eq.(\ref{31_Mf1_x0to0}), we obtain the following expression:
\begin{eqnarray}
\label{31_Mf2_x0to0}
&F_{I\hspace{-2.5pt}I}(t;x_0 \Rightarrow0) = F_I(t;x_0\Rightarrow0) \nonumber 
- \int_{0}^{t} d\tau_1 F_I(\tau_1;x_0\Rightarrow L) F_I(t-\tau_1;L\Rightarrow0) \nonumber \\
&+ \int_{0}^{t} d\tau_1 \int_{0}^{\tau_1} d\tau_2 F_{I\hspace{-2.5pt}I}(\tau_2;x_0\Rightarrow 0)
   F_I(\tau_1-\tau_2;0\Rightarrow L) F_I(t-\tau_1;L\Rightarrow0).
\end{eqnarray}
To interpret Eq.(\ref{31_Mf2_x0to0}), let us tentatively neglect the last term in Eq.(\ref{31_Mf2_x0to0}).
Then, the second term represents the FPT probability associated with {\it wrong} trajectories that a random walker reaches $L$ earlier than $0$.
At first sight, one may wonder if we successfully eliminate {\it wrong} trajectories using only $F_I$.
However, this results in an underestimation of $F_{I\hspace{-2.5pt}I}$ since the second term includes the contribution of excess trajectories, in which a random walker touches boundary $0$ at first, then boundary $L$, and finally boundary $0$ again at time $t$ (Fig.\ref{fig_trajectory}(b)).
The correction due to such trajectories is expressed by the last term, which again includes $F_{I\hspace{-2.5pt}I}$.

By repeating the filtering operation $N$ times, we obtain the following equation:
\begin{eqnarray}
\label{31_F2_t_f1+f2}
  F_{I\hspace{-2.5pt}I}&(t;x_0\Rightarrow0) = \sum_{n=0}^{N-1} (-1)^n f_I^{(n)}(t;x_0\Rightarrow0)
+ (-1)^{N} f_{I\hspace{-2.5pt}I}^{(N)}(t;x_0\Rightarrow0),
\end{eqnarray}
where $f_{I}^{(n)}$ and $f_{I,{I\hspace{-2.5pt}I}}^{(n)}$ is defined recursively as follows: 
\begin{eqnarray}
\label{31_f12_t_0}
  f_{I,{I\hspace{-2.5pt}I}}^{(0)}(t;x_0\Rightarrow B) = F_{I,{I\hspace{-2.5pt}I}}(t;x_0\Rightarrow B)
\qquad (n=0),
\end{eqnarray}
\begin{eqnarray}
\label{31_f12_t_n}
  f_{I,{I\hspace{-2.5pt}I}}^{(n)}(t;x_0\Rightarrow B_1) = 
   \int_{0}^{t} d\tau_n f_{I,{I\hspace{-2.5pt}I}}^{(n-1)}(\tau_n,x_0 \Rightarrow B_2) F_I(t-\tau_n,B_2 \Rightarrow B_1),
   \, (n \geq 1),
\end{eqnarray}
where $B = 0$ or $L$ in Eq.(\ref{31_f12_t_0}) and $(B_1,B_2)=(0,L)$ or $(L,0)$ in Eq.(\ref{31_f12_t_n}). 
The subscripts of $f^{(n)}_{I,{I\hspace{-2.5pt}I}}$ on the left hand in Eqs.(\ref{31_f12_t_0}) and (\ref{31_f12_t_n}) correspond to those on the right hand, respectively, in order.
For convenience, we put subscripts of $\tau$ in Eq.(\ref{31_f12_t_n}) in the reverse order of Eq.(\ref{31_Mf2_x0to0}).
In the Laplace domain, Eq.(\ref{31_F2_t_f1+f2}) is 
\begin{eqnarray}
\label{31_F2_s_f1+f2}
  \tilde{F}_{I\hspace{-2.5pt}I}&(s;x_0\Rightarrow0) = \sum_{n=0}^{N-1} (-1)^n \tilde{f}_I^{(n)}(s;x_0\Rightarrow0)
+ (-1)^{N} \tilde{f}_{I\hspace{-2.5pt}I}^{(N)}(s;x_0\Rightarrow0).
\end{eqnarray}
We note that $f_{I,I\hspace{-2.5pt}I}^{(n)}$ can be expressed as a simple product in the Laplace domain
\begin{eqnarray}
\label{31_f12_s_Lap}
\tilde{f}_{I,I\hspace{-2.5pt}I}^{(n)}(s;x_0\Rightarrow 0) \nonumber   = 
\begin{cases}
         \tilde{F}_{I,I\hspace{-2.5pt}I}(s;x_0\Rightarrow 0) \left[
         \tilde{F_I}(s;0\Rightarrow L) \tilde{F_I}(s;L\Rightarrow 0) \right] ^{n/2} & n $ is even$ \\
         \tilde{F}_{I,I\hspace{-2.5pt}I}(s;x_0\Rightarrow L)\tilde{F_I}(s;L\Rightarrow 0) 
         \left[ \tilde{F_I}(s;0\Rightarrow L) \tilde{F_I}(s;L\Rightarrow 0) \right] ^{(n-1)/2} & n$ is odd.$ \\
\end{cases}
\end{eqnarray}
When $N$ is even number, the last term in Eq.(\ref{31_F2_s_f1+f2}) can be written as a function of $\tilde{F}_{I\hspace{-2.5pt}I}(s;x_0\Rightarrow0)$ as follows:
\begin{eqnarray}
\label{31_f2_s_last}
  (-1)^{N} \tilde{f}_{I\hspace{-2.5pt}I}^{(N)}(s;x_0\Rightarrow0)
  = \tilde{F}_{I\hspace{-2.5pt}I}&(s;x_0\Rightarrow0)  
  \left[ \tilde{F_I}(s;0\Rightarrow L) \tilde{F_I}(s;L\Rightarrow 0) \right] ^{N/2}.
\end{eqnarray}
Therefore, after elementary algebraic calculations, Eq.(\ref{31_F2_s_f1+f2}) can be rewritten as follows:
\begin{eqnarray}
\label{31_F2_s_N}
  \tilde{F}_{I\hspace{-2.5pt}I}&(s;x_0\Rightarrow0) = 
    \frac{\sum_{n=0}^{N-1} (-1)^n \tilde{f}_I^{(n)}(s;x_0\Rightarrow0)}
         {1 - \left[ \tilde{F_I}(s;0\Rightarrow L) \tilde{F_I}(s;L\Rightarrow 0) \right] ^{N/2}}.
\end{eqnarray}
This formula represents the solution with two absorbing boundaries $F_{I\hspace{-2.5pt}I}$ using only solutions with one absorbing boundary $F_I$.
Two of our main results are obtained from Eq.(\ref{31_F2_s_N}) as follows:
\begin{itemize}
  \item The simplest solution form is obtained by setting $N=2$,
    \begin{eqnarray}
    \label{31_F2_s_2}
      \tilde{F}_{I\hspace{-2.5pt}I}&(s;x_0\Rightarrow0) = 
        \frac{\tilde{F_I}(s;x_0\Rightarrow 0)-\tilde{F_I}(s;x_0\Rightarrow L)\tilde{F_I}(s;L\Rightarrow 0)}
         {1 - \tilde{F_I}(s;0\Rightarrow L) \tilde{F_I}(s;L\Rightarrow 0) }.
    \end{eqnarray}
  \item Another simple expression is obtained, if $|\tilde{F_I}(s;0\Rightarrow L) \tilde{F_I}(s;L\Rightarrow 0)|<1$,
        taking limit $N \to \infty$ and performing the inverse Laplace transform of Eq.(\ref{31_F2_s_N}),
    \begin{eqnarray}
    \label{31_F2_t_inf}
      F_{I\hspace{-2.5pt}I}(t;x_0\Rightarrow0) = 
       \lim\limits_{N\to\infty} \sum_{n=0}^{N-1} (-1)^n f_I^{(n)}(t;x_0\Rightarrow0).
    \end{eqnarray}
        The convergence of the summation in Eq.(\ref{31_F2_t_inf}) can be checked from a ratio test 
        $\lim\limits_{n\to\infty} \left| \frac{f_I^{(n+1)}}{f_I^{(n)}} \right| < 1$.
\end{itemize}

\section{Applications}
We provide concrete forms of two expressions Eqs.(\ref{31_F2_s_2}) and (\ref{31_F2_t_inf}) for some examples.
\subsection{Free diffusion}
In free diffusion, transition probability in infinite space is 
\begin{equation}
\label{41_traP}
  P(x,t|x_0,t_0) = \frac{1}{\sqrt{4\pi D(t-t_0)}} \exp{\left[-\frac{(x-x_0)^2}{4D(t-t_0)}\right]}
\end{equation}
\cite{FPequ}.
Using Eqs.(\ref{22_conv_Lap}), (\ref{22_conv_ILap}), and (\ref{41_traP}), we obtain $\tilde{F}_I$ and $F_I$ as follows:
\begin{equation}
\label{41_F1_s}
\tilde{F_I}(s;A\Rightarrow B) = \exp(-\sqrt{2\tau_{A}s}) ,
\end{equation}
\begin{equation}
\label{41_F1_t}
F_I(t;A \Rightarrow B) = \frac{1}{\sqrt{2\pi}\tau_{A}} 
   \exp{\left[ -\frac{\tau_{A}} {2t} \right]}
   \left(\frac{t}{\tau_{A}} \right)^{-\frac{3}{2}},
\end{equation}
where $\tau_A = \frac{(B-A)^2}{2D} $.
First of all, we derive $F_{I\hspace{-2.5pt}I}(t;x_0 \Rightarrow 0)$ from Eq.(\ref{31_F2_s_N}).
By substituting Eq.(\ref{41_F1_s}) into Eq.(\ref{31_F2_s_N}), we find
\begin{eqnarray}
\label{41_F2_s_N}
  \tilde{F}_{I\hspace{-2.5pt}I}&(s;x_0\Rightarrow0) = 
    \frac{\sum_{n=0}^{N/2-1} \sinh{\left[ \sqrt{\frac{s}{D}} (\frac{NL}{2} - x_0 - 2Ln)  \right]}}
         { \sinh{\left[ \sqrt{\frac{s}{D}} \frac{NL}{2} \right]} }.
\end{eqnarray}
%
%%%%%%%%%%%%%%%%%%%%%%%%%%%%%%%%%%%%%%%%%%%%%%%%%%%%%%%%%%%%%%%%%%%%%%%%%%%%%%%%%%%%%%%%%%%%%%%%
\begin{figure}[h]
\centering
\begin{tabular}{cc}
  \begin{minipage}[b]{70mm}
      \centering
      \includegraphics[width=70mm]{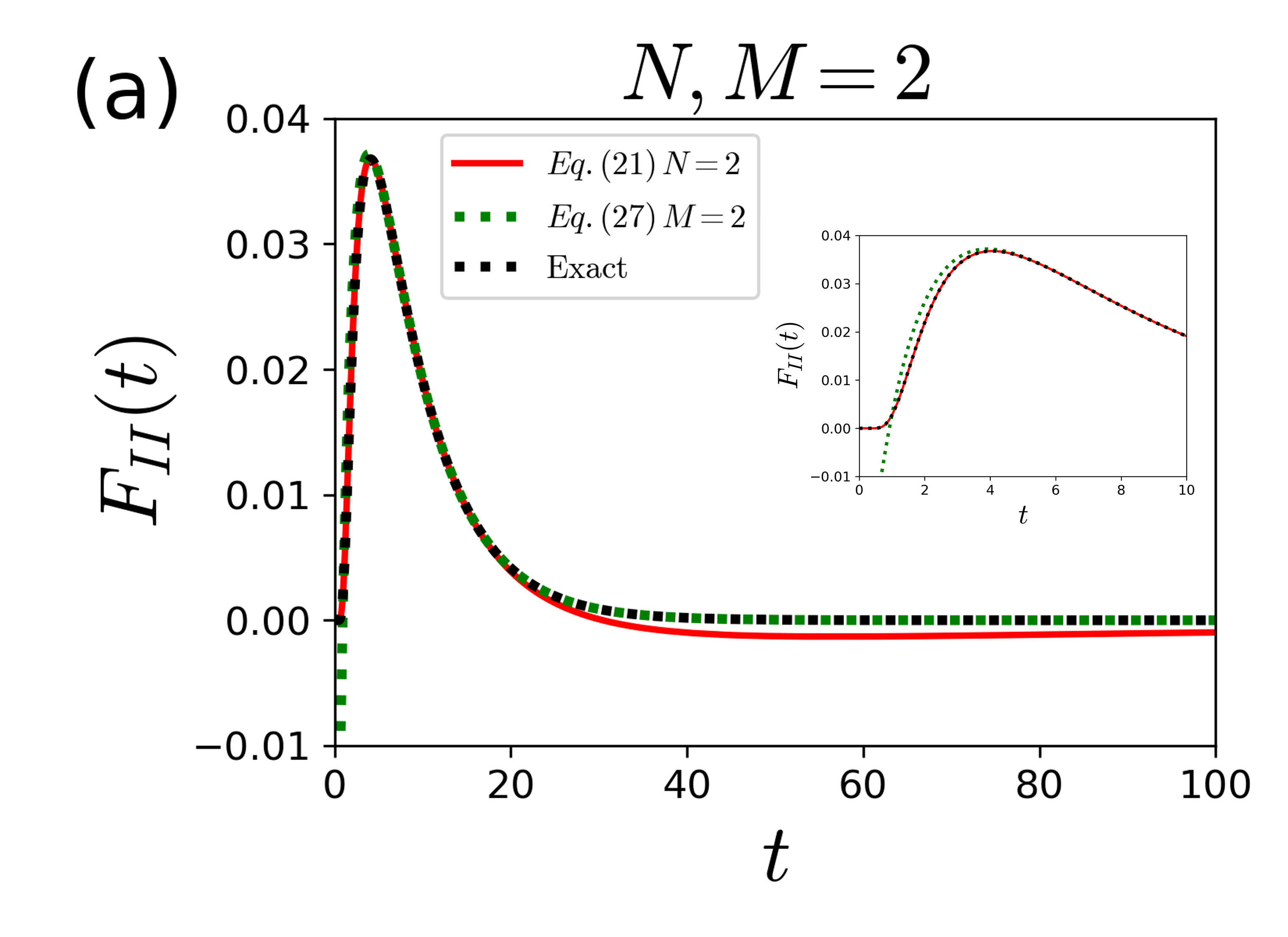}
      %\subcaption{}
      %\label{fou}
  \end{minipage} &
  \begin{minipage}[b]{70mm}
      \centering
      \includegraphics[width=70mm]{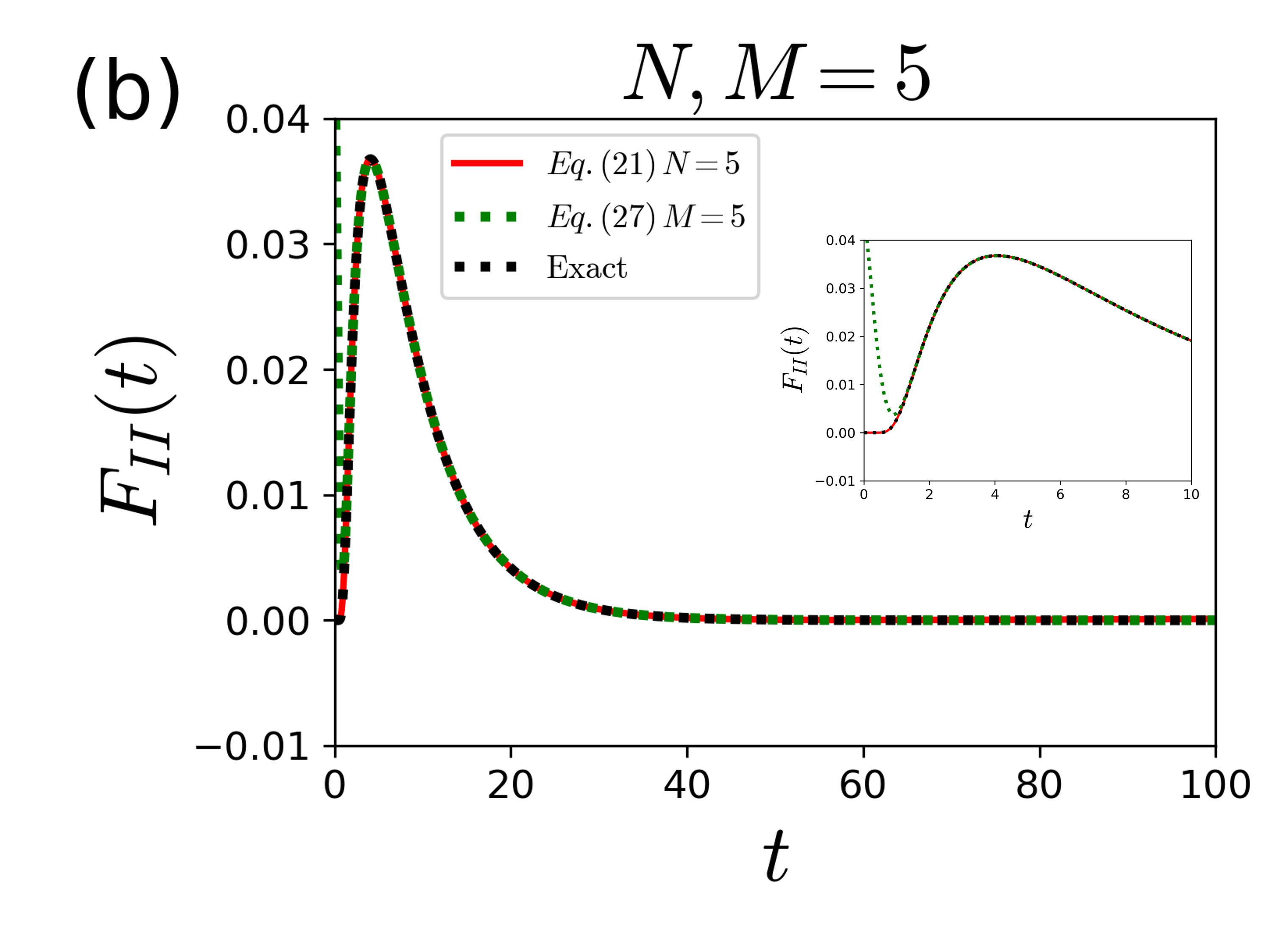}
      %\subcaption{}
      %\label{fig:b}
  \end{minipage} \\
  \begin{minipage}[b]{0.0mm}
      \centering
      \includegraphics[width=70mm]{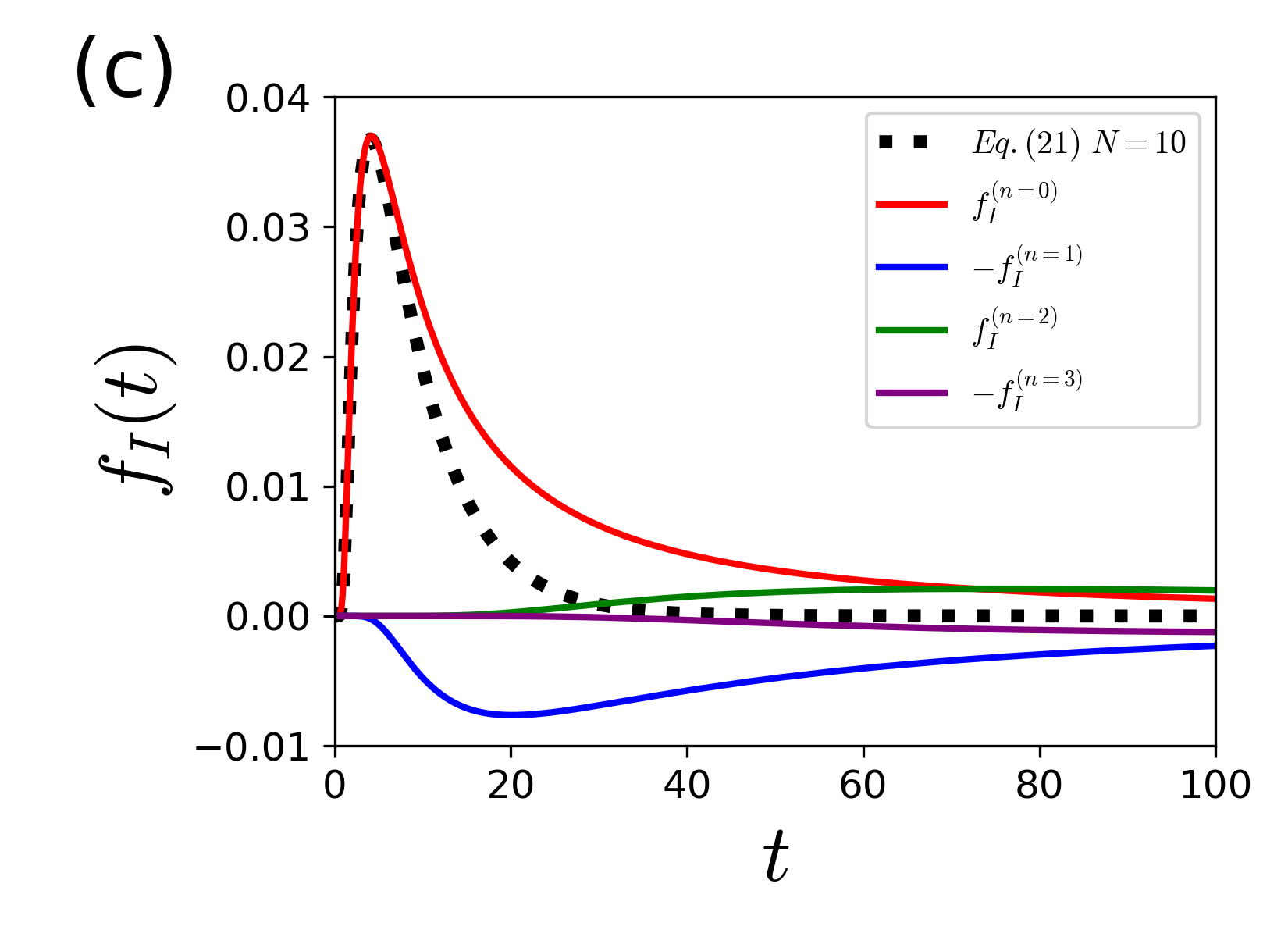}
      %\subcaption{}
      %\label{fig:c}
  \end{minipage} 
\end{tabular}
\caption{Results for FPT probability $F_{I\hspace{-2.5pt}I}(t;x_0\Rightarrow0)$ in free diffusion, where the initial position $x_0=5$, the location of right absorbing boundary $L=8$.
(a)(b) Comparison of Eq.(\ref{31_F2_t_inf}) (solid lines) and the method of eigenfunction expansion (EE method) Eq.(\ref{41_F2_t_2}) (dotted line) for FPT distribution $F_{I\hspace{-2.5pt}I}(t)$
(Inset shows short-time region).
The exact solution is obtained from the EE method with large enough $M$ (here $M=30$).
(c) Individual $f_I^{(n)}$ obtained from Eq.(\ref{41_f1_t}).
The functions causing underestimation are represented as negative.
 }
\label{41_fig_F}
\end{figure}
%%%%%%%%%%%%%%%%%%%%%%%%%%%%%%%%%%%%%%%%%%%%%%%%%%%%%%%%%%%%%%%%%%%%%%%%%%%%%%%%%%%%%%%%%%%%%%%%
By performing inverse Laplace transform of Eq.(\ref{41_F2_s_N}) (see Appendix B), we obtain
\begin{eqnarray}
\label{41_F2_t_N}
  F_{I\hspace{-2.5pt}I}&(t;x_0\Rightarrow0) = 
    \frac{8D\pi}{N^2L^2} \sum_{m=1}^{\infty} m e^{-(\frac{2m\pi}{NL})^2 Dt}
     \sum_{n=0}^{N/2-1} \sin \left[ \left(\frac{2x_0}{L}+4n\right) \frac{m\pi}{N} \right] .
\end{eqnarray}
Setting $N=2$ or directly from Eq.(\ref{31_F2_s_2}), we find
\begin{eqnarray}
\label{41_F2_t_2}
  F_{I\hspace{-2.5pt}I}&(t;x_0\Rightarrow0) = 
    \lim\limits_{M\to\infty} \frac{2D\pi}{L^2} \sum_{m=1}^{M} m \sin \left(\frac{m\pi}{L}x_0 \right) e^{-t/\tau^E_m},
\end{eqnarray}
where $\tau^{E}_m = \frac{L^2}{D} (\frac{1}{m\pi})^2 $.
Remarkably, Eq.(\ref{41_F2_t_2}) agrees with the result from the method of eigenfunction expansion (EE method) [1].
Note that the origin of summation over $m$ in Eq.(\ref{41_F2_t_2}) (also in Eq.(\ref{41_F2_t_N})) is not the repeated filtration operations.
Rather, it comes from the contribution of the infinite number of zeros in the denominator of Eq.(\ref{41_F2_s_N}), which leads to poles in the inverse Laplace transformation (see Appendix B).
In addition, Eqs.(\ref{41_F2_t_N}) and (\ref{41_F2_t_2}) are indeed equivalent,
i.e., a change of variables $m \to \frac{N}{2}m$ in the former transforms to the latter.

Next, we look at an alternative expression of $F_{I\hspace{-2.5pt}I}(t;x_0 \Rightarrow 0)$ given by Eq.(21),
where we use Eq.(\ref{41_F1_s}) for free diffusion, together with Eq.(\ref{31_f12_s_Lap}) to obtain
\begin{eqnarray}
\label{41_f1_t}
f_I^{(n)}(t;x_0\Rightarrow 0) = 
\begin{cases}
  F_I(t;Ln + x_0 \Rightarrow 0)& n$ is even$  \\
  F_I(t;L(n+1)-x_0 \Rightarrow 0) & n$ is odd.$ \\
\end{cases}       
\end{eqnarray}
Hence, this second expression from Eq.(\ref{31_F2_t_inf}) reduces to the result obtained by the method of image [1].

Although both Eqs.(\ref{31_F2_t_inf}) and (\ref{41_F2_t_2}) are exact in large $N$ or $M$ limits, it is interesting to see the performance of approximation truncated at finite $N,M$.
For the first expression Eq.(\ref{41_F2_t_2}), $\tau^{E}_m$ is the relaxation time of $m$-th mode.
Thus, to describe FPT distribution in the time scale $\sim t$, taking the terms up to $m$-th mode that satisfies $\tau^{E}_m \sim t$ would be necessary.
Similarly, for the second expression Eq.(\ref{31_F2_t_inf}), we introduce a characteristic time $t^*_{(n)}$ in $n$-th filtration operation from the maximum of $f_I^{(n)}$, i.e., $ \frac{\partial}{\partial t} f_I^{(n)}(t)|_{t=t^*_{(n)}} = 0 $.
In free diffusion, we find 
\begin{eqnarray}
\label{41_tstar}
t^*_{(n)} = 
\begin{cases}
  \frac{(Ln+x_0)^2}{6D} & n$ is even$  \\
  \frac{(L(n+1)-x_0)^2}{6D} & n$ is odd,$ \\
\end{cases}  
\end{eqnarray}
which corresponds to characteristic travel time, starting from $x_0$ to the absorbing boundary at $x=0$ in a one-absorbing setting, with $n$ times absorption in a two-boundary setting (see Fig.1).
As illustrated in Fig.\ref{41_fig_F} (a),(b), the short-time regime (initial sharp rise and following peak) is accurately described already at $N=2$ $(t^*_{(1)}\sim20)$, while the long-time regime requires larger $N$.
In the example shown here, the truncation at $N=5$ $(t^*_{(4)}\sim230)$ provides the FPT distribution indistinguishable from the exact result in the shown time regime.
Fig.\ref{41_fig_F} (c) shows individual contributions $f_I^{(n)}(t)$, which represent the effect of successive filtration operations to construct progressively accurate $F_{I\hspace{-2.5pt}I}(t)$.
%%%%%%%%%%%%%%%%%%%%%%%%%%%%%%%%%%%%%%%%%%%%%%%%%%%%%%%%%%%%%%%%%%%%%%%%%%%%%%%%%%%%%%%%%%%%%
%%%%%%%%%%%%%%%%%%%%%%%%%%%%%%%%%%%%%%%%%%%%%%%%%%%%%%%%%%%%%%%%%%%%%%%%%%%%%%%%%%%%%%%%%%%%%
%%%%%%%%%%%%%%%%%%%%%%%%%%%%%%%%%%%%%%%%%%%%%%%%%%%%%%%%%%%%%%%%%%%%%%%%%%%%%%%%%%%%%%%%%%%%%

% 4-2
\subsection{Biased diffusion}
Next, let us discuss Brownian motion under the influence of linear potential $U(x)=-\alpha x$.
Transition probability in infinite space is 
\begin{equation}
\label{42_traP}
  P(x,t|x_0,t_0) = \frac{1}{\sqrt{4\pi D(t-t_0)}}
    \exp{\left[-\frac{(x-x_0-{v(t-t_0)})^2}{4D(t-t_0)}\right]}
\end{equation}
\cite{FPequ}, where $v=\frac{\alpha}{\gamma}$.
Using Eqs.(\ref{22_conv_Lap}), (\ref{22_conv_ILap}) and (\ref{42_traP}), 
we find
\begin{equation}
\label{42_F1_s}
  \tilde{F_I}(s;A \Rightarrow B) = \exp{\left[ -\frac{\sqrt{4Ds + v^2}|B-A| - (B-A)v}{2D} \right] },
\end{equation}
\begin{equation}
\label{42_F1_t}
  F_I(t;A \Rightarrow B) = \frac{|A-B|}{\sqrt{4 \pi D t^3}} 
   \exp{\left[ -\frac{(A-B+vt)^2} {4Dt} \right]}.
\end{equation}

From Eqs.(\ref{31_F2_s_2}) and (\ref{42_F1_s}), we obtain the first expression
\begin{eqnarray}
\label{42_F2_s_2}
  \tilde{F}_{I\hspace{-2.5pt}I}&(s;x_0\Rightarrow0) = 
    \exp{\left[ \frac{-x_0 v}{2D} \right]}
    \frac{\sinh{\left[ \frac{\sqrt{4Ds+v^2} (L-x_0)}{2D} \right]}}
         {\sinh{\left[ \frac{\sqrt{4Ds+v^2} L}{2D} \right]}}.
\end{eqnarray}
By performing the inverse Laplace transform of Eq.(\ref{42_F2_s_2}) in the same procedure as the free diffusion case [see Appendix B],
we obtain 
\begin{eqnarray}
\label{42_F2_t_2}
  F_{I\hspace{-2.5pt}I}&(t;x_0\Rightarrow0) = 
    \lim\limits_{M\to\infty} \frac{2D\pi}{L^2} \exp{\left[ \frac{x_0 v}{2D} \right]}
    \sum_{m=1}^{M} m \sin \left(\frac{m\pi}{L}x_0 \right) e^{-\left[(\frac{m\pi}{L})^2 D + \frac{v^2}{4D}\right]t},
\end{eqnarray}
where $\tau^{E}_m = \frac{L^2}{D} \frac{1}{(m\pi)^2 + \left(\frac{Lv}{2D}\right)^2} $.
Again, Eq.(\ref{42_F2_t_2}) agrees with the EE method [1].

For the second expression Eq.(\ref{31_F2_t_inf}), 
we use Eq.(\ref{42_F1_s}) for biased diffusion, together with Eq.(\ref{31_f12_s_Lap}) to find
\begin{eqnarray}
\label{42_f1_t}
f_I^{(n)}(t;x_0\Rightarrow 0) = 
\begin{cases}
  F_I(t;Ln + x_0 \Rightarrow 0) e^{\frac{Lv}{2D}n} & $n$ is even  \\
  F_I(t;0 \Rightarrow L(n+1)- x_0 ) e^{-\frac{Lv}{2D}(n+1)} & $n$ is odd. \\
\end{cases}  
\end{eqnarray}
%
%%%%%%%%%%%%%%%%%%%%%%%%%%%%%%%%%%%%%%%%%%%%%%%%%%%%%%%%%%%%%%%%%%%%%%%%%%%%%%%%%%%%%%%%%%%%%
\begin{figure}[h]
\centering
\begin{tabular}{cc}
  \begin{minipage}[b]{70mm}
      \centering
      \includegraphics[width=70mm]{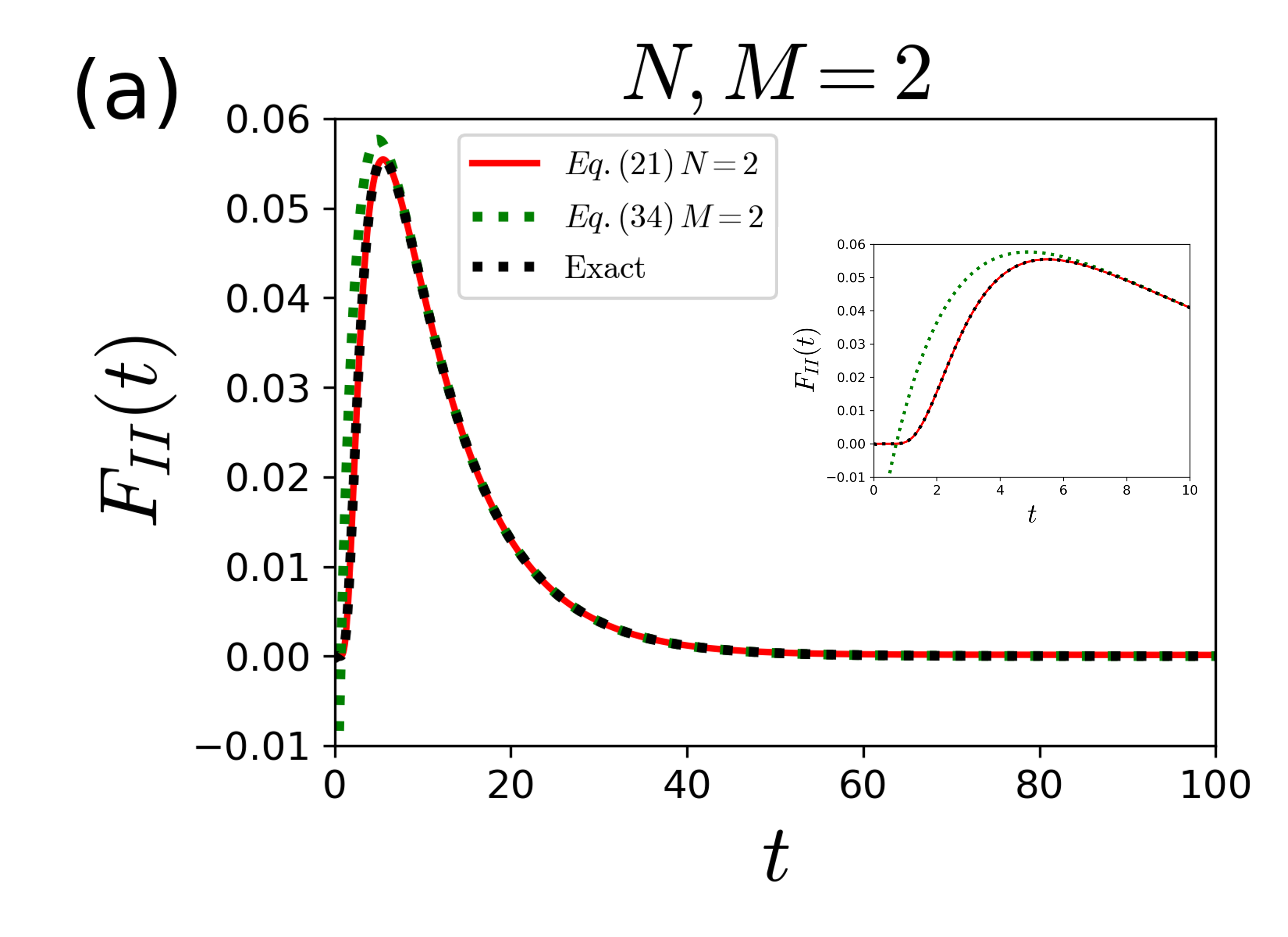}
      %\subcaption{}
      %\label{fou}
  \end{minipage} &
  \begin{minipage}[b]{70mm}
      \centering
      \includegraphics[width=70mm]{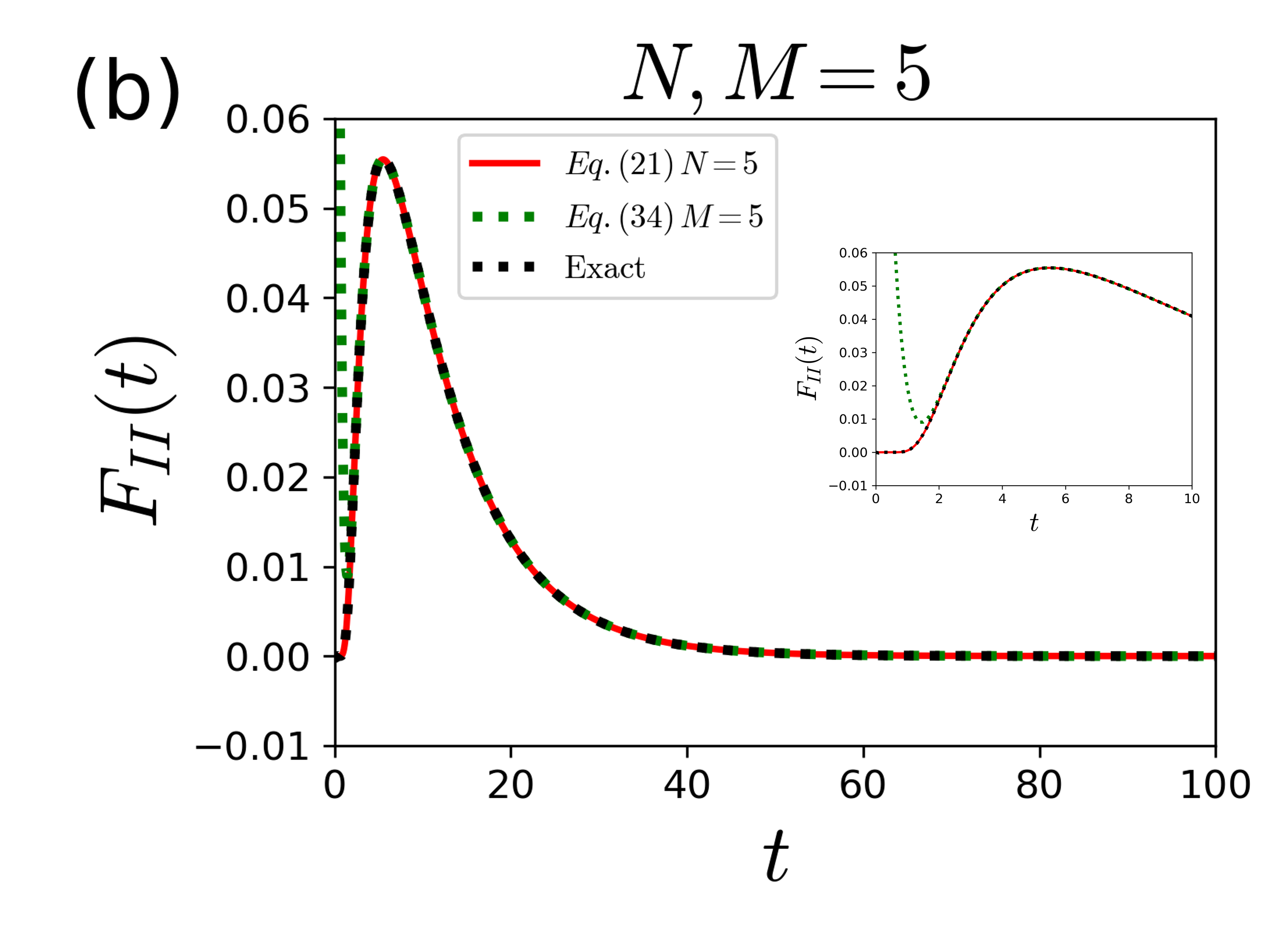}
      %\subcaption{}
      %\label{fig:b}
  \end{minipage} \\
  \begin{minipage}[b]{0.0mm}
      \centering
      \includegraphics[width=70mm]{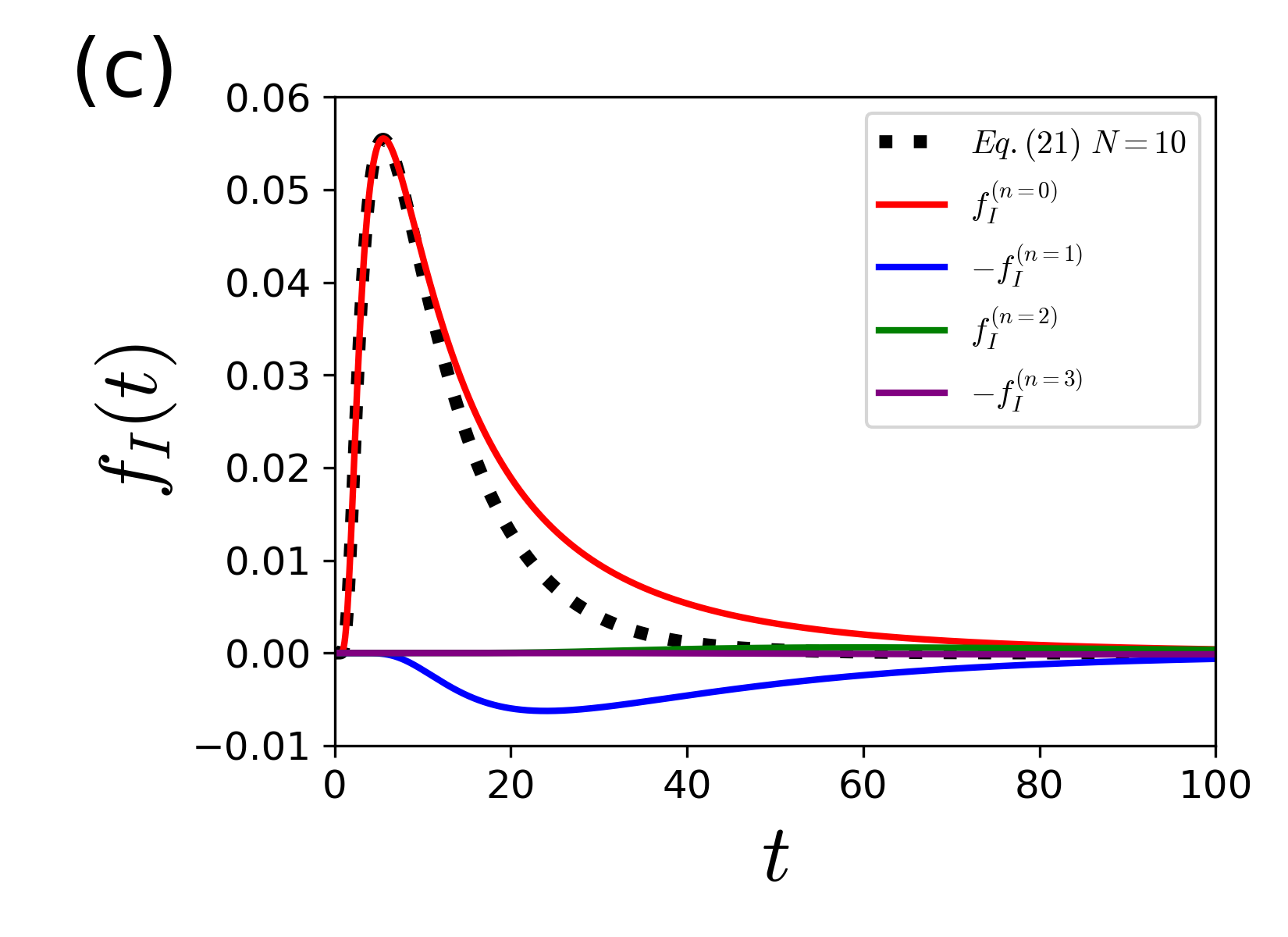}
      %\subcaption{}
      %\label{fig:c}
  \end{minipage} 
\end{tabular}
\caption{Results for FPT probability $F_{I\hspace{-2.5pt}I}(t;x_0\Rightarrow0)$ in biased diffusion, where the initial position $x_0=6$, the location of right absorbing boundary $L=10$, and the slope of the potential $\alpha=-0.3$.
(a)(b) Comparison of Eq.(\ref{31_F2_t_inf}) (solid lines) and the EE method Eq.(\ref{42_F2_t_2}) (dotted line) for FPT distribution $F_{I\hspace{-2.5pt}I}(t;x_0 \Rightarrow 0)$ (Inset shows-short time region).
The exact solution is obtained from the EE method with a large enough $M$ (here, $M=30$).
(c) Individual $f_I^{(n)}$ obtained from Eq.(\ref{42_f1_t}).
The functions causing underestimation are represented as negative.}
\label{42_fig_F}
\end{figure}
%%%%%%%%%%%%%%%%%%%%%%%%%%%%%%%%%%%%%%%%%%%%%%%%%%%%%%%%%%%%%%%%%%%%%%%%%%%%%%%%%%%%%%%%%%%%%%%%
Similar to the free diffusion case, our second form reduces to that obtained by the method of image,
appropriately modified with multiplication factor $e^{\frac{Lv}{D}n}$ to incorporate the bias effect \cite{guide}.
For the characteristic time,
\begin{eqnarray}
\label{42_tstar}
t^*_{(n)} = 
\begin{cases}
  \frac{-3D + \sqrt{9D^2 + v^2(Ln+x_0)^2}}{v^2} & $n$ is even  \\
  \frac{-3D + \sqrt{9D^2 + v^2(Ln+L-x_0)^2}}{v^2} & $n$ is odd. \\
\end{cases}
\end{eqnarray}
The performance of finite $N$ truncation approximation is shown in Fig.\ref{42_fig_F}.
%%%%%%%%%%%%%%%%%%%%%%%%%%%%%%%%%%%%%%%%%%%%%%%%%%%%%%%%%%%%%%%%%%%%%%%%%%%%%%%%%%%%%%%%%%%%%
%%%%%%%%%%%%%%%%%%%%%%%%%%%%%%%%%%%%%%%%%%%%%%%%%%%%%%%%%%%%%%%%%%%%%%%%%%%%%%%%%%%%%%%%%%%%%
% 4-3
\subsection{Harmonic Potential}
%%%%%%%%%%%%%%%%%%%%%%%%%%%%%%%%%%%%%%%%%%%%%%%%%%%%%%%%%%%%%%%%%%%%%%%%%%%%%%%%%%%%%%%%%%%%%
\begin{figure}[h]
\centering
\begin{tabular}{cc}
  \begin{minipage}[b]{70mm}
      \centering
      \includegraphics[width=70mm]{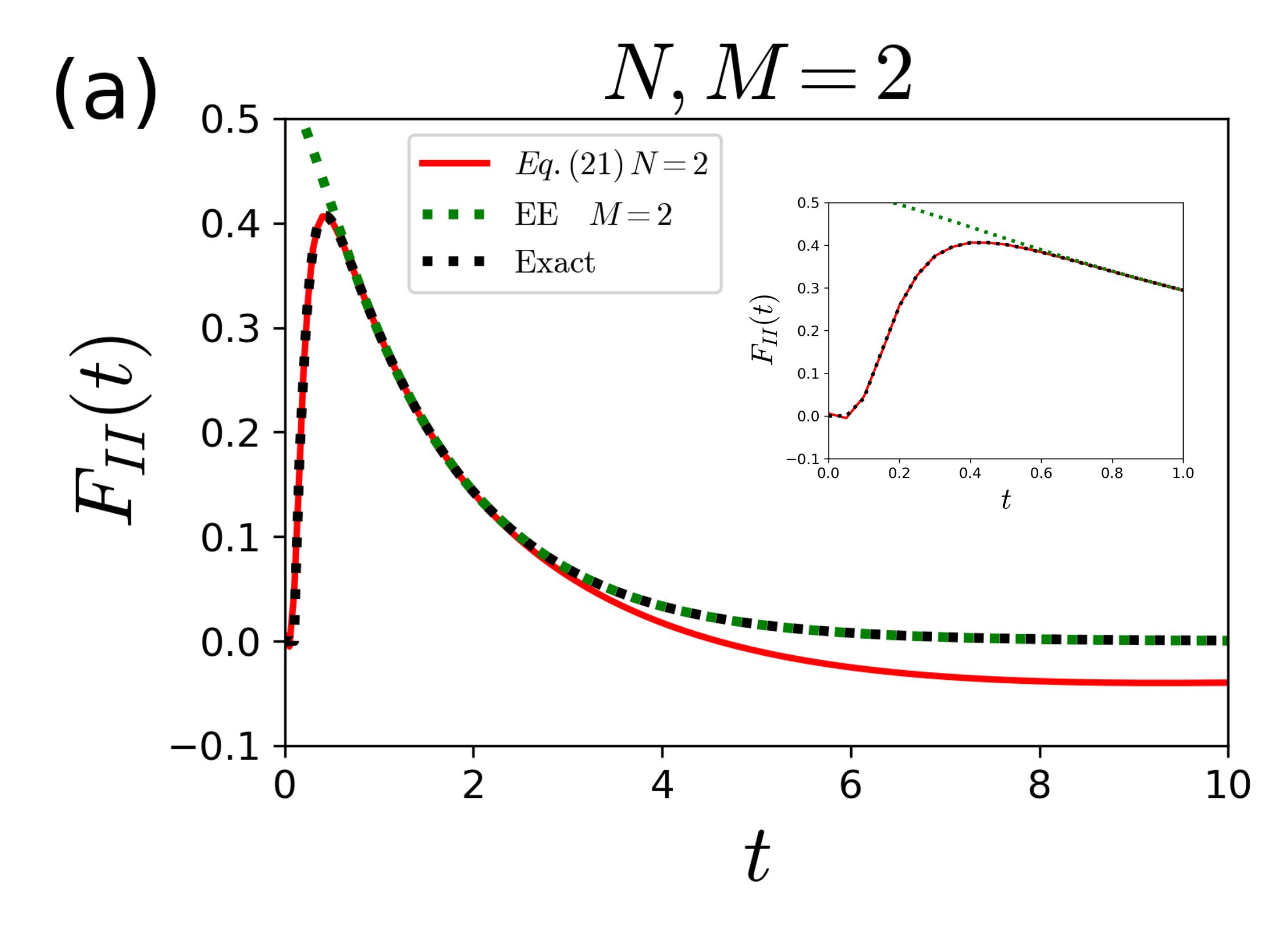}
      %\subcaption{}
      %\label{fou}
  \end{minipage} &
  \begin{minipage}[b]{70mm}
      \centering
      \includegraphics[width=70mm]{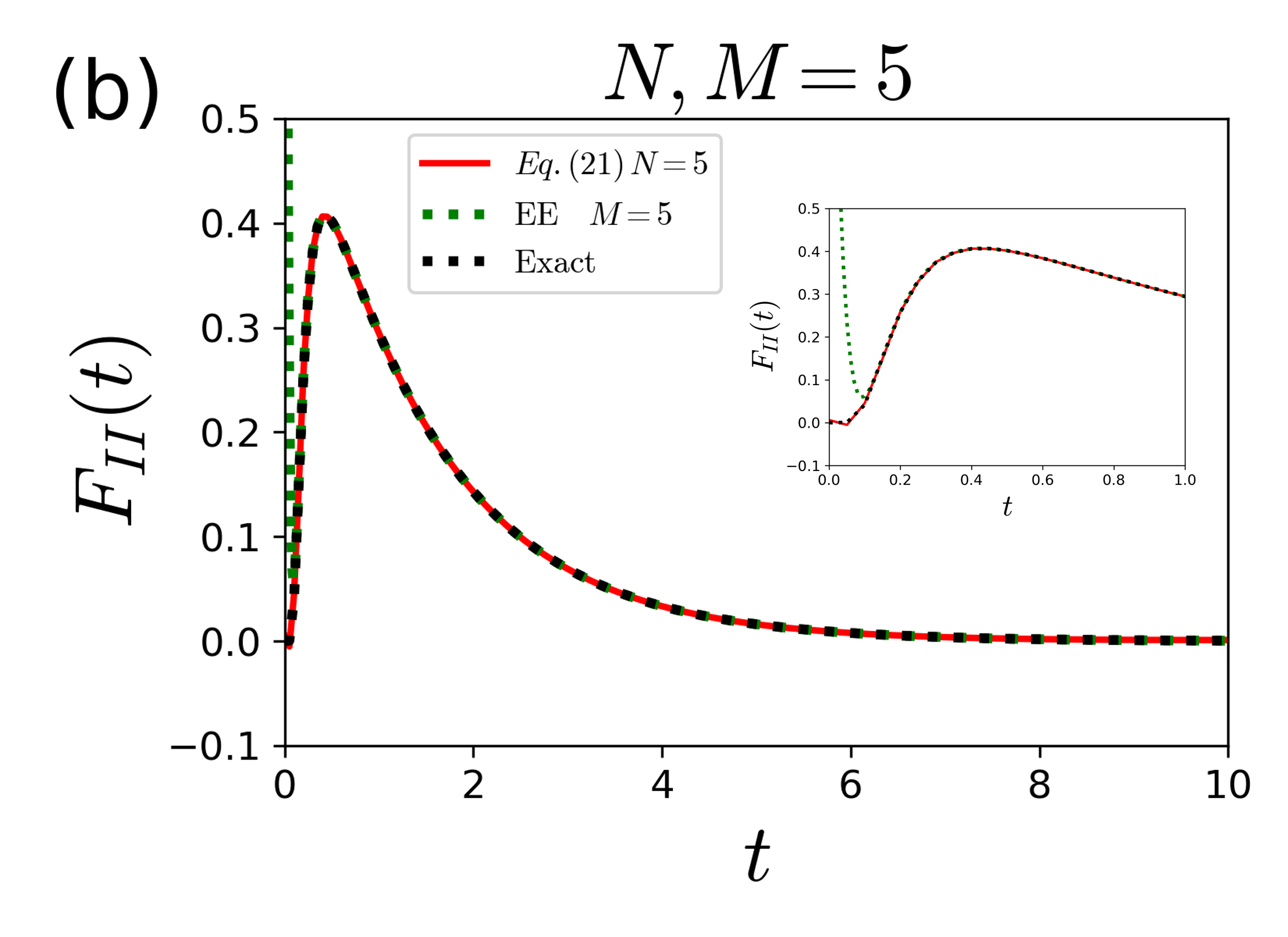}
      %\subcaption{}
      %\label{fig:b}
  \end{minipage} \\
  \begin{minipage}[b]{70mm}
      \centering
      \includegraphics[width=70mm]{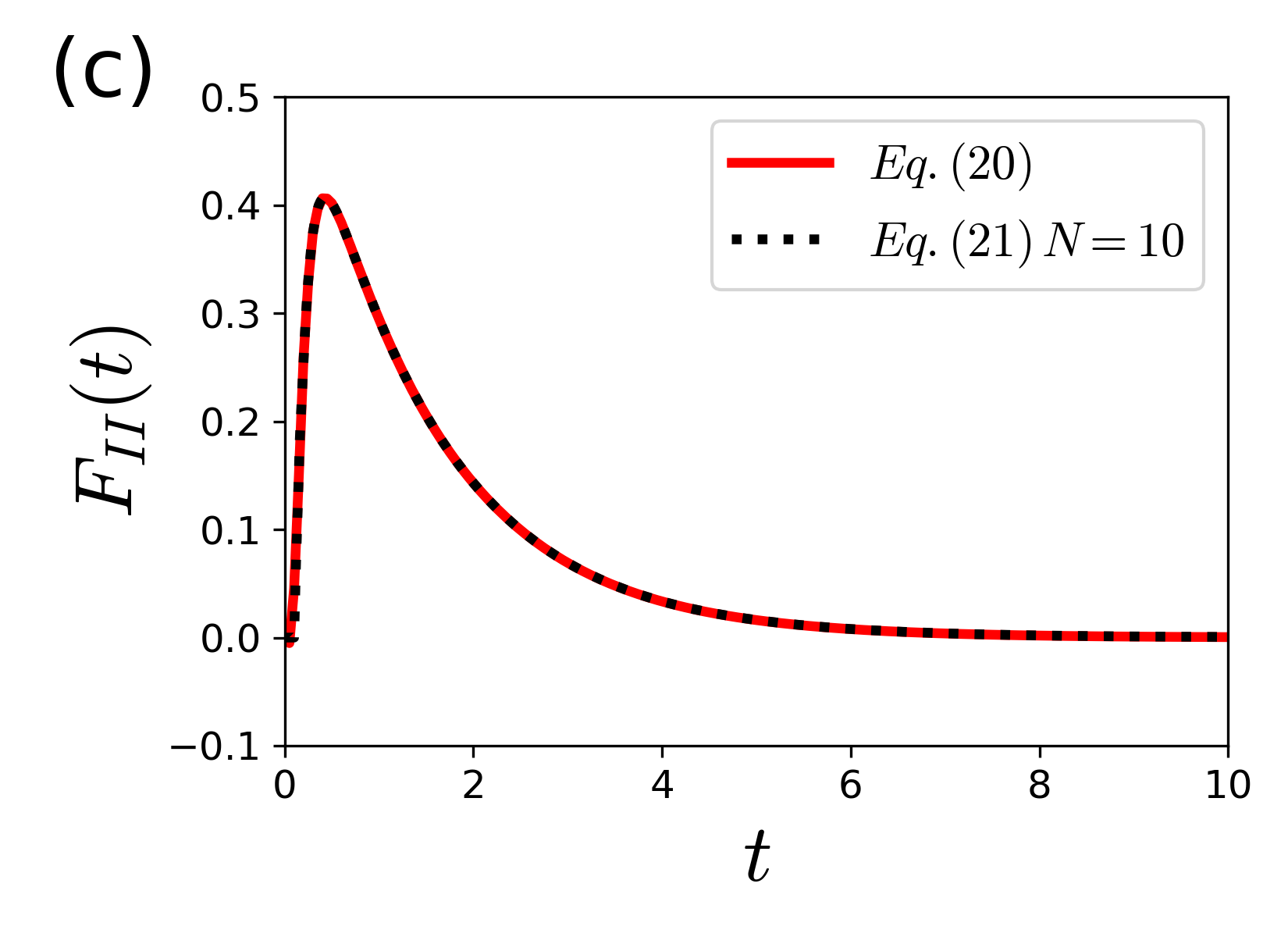}
      %\subcaption{}
      %\label{fig:c}
  \end{minipage} &
  \begin{minipage}[b]{70mm}
      \centering
      \includegraphics[width=70mm]{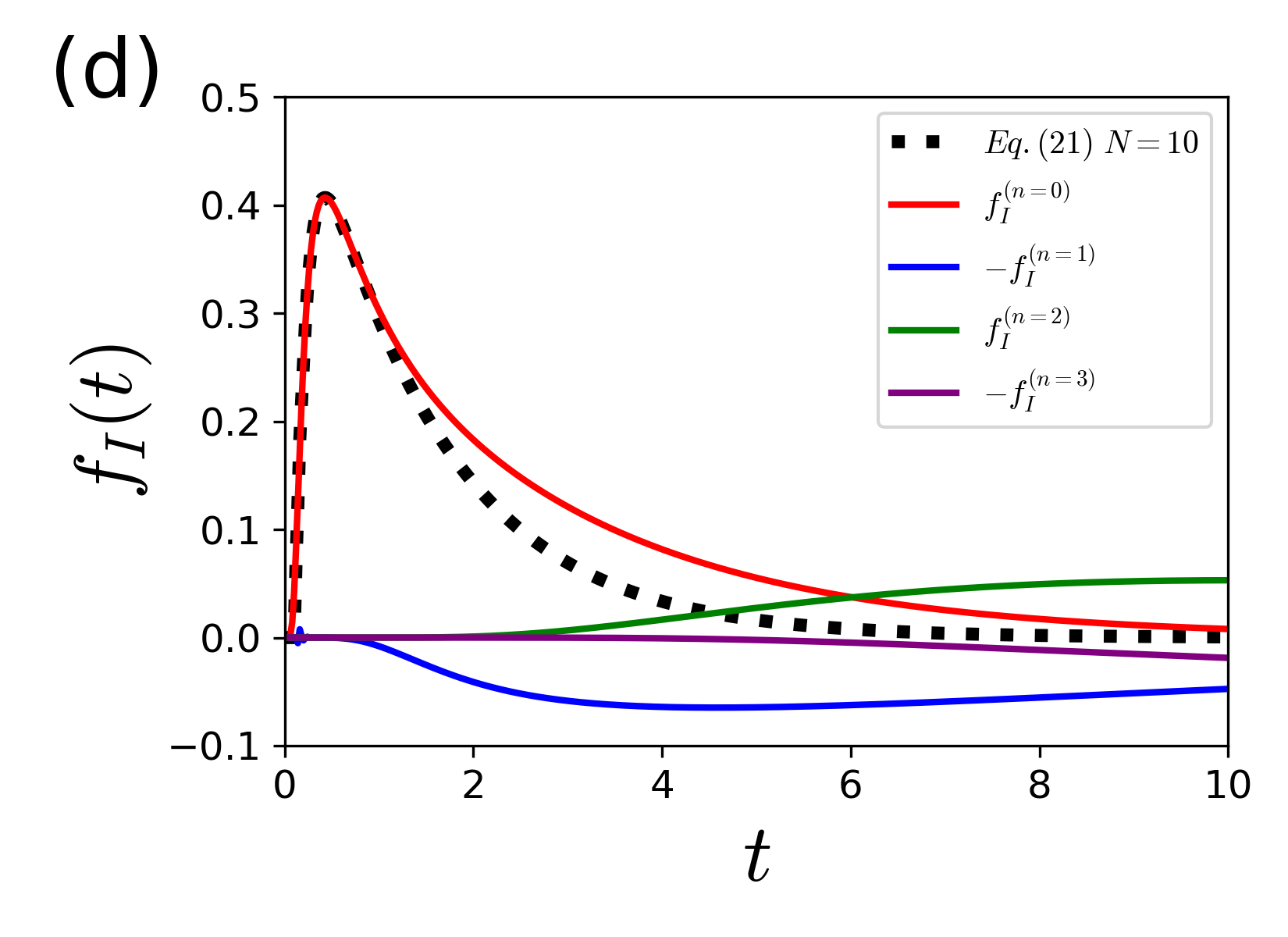}
      %\subcaption{}
      %\label{fig:c}
  \end{minipage} 
\end{tabular}
\caption{Results for FPT probability $F_{I\hspace{-2.5pt}I}(t;x_0\Rightarrow0)$ in harmonic potential, where $a=1$, the initial position $x_0=1.5$, the location of right absorbing boundary $L=3$, spring constant $k=1$.
(a)(b) Comparison of Eq.(\ref{31_F2_t_inf}) (solid lines) and the EE method (dotted line) for FPT distribution $F_{I\hspace{-2.5pt}I}(t;x_0 \Rightarrow 0)$ (Inset shows-short time region).
The exact solution is obtained from the EE method with a large enough $M$ (here, $M=30$).
(c) Comparison of inverse Laplace transform of Eq.(\ref{31_F2_s_2}) (dotted line) and Eq.(\ref{31_F2_t_inf}) with $N=10$ (solid lines) for FPT distribution $F_{I\hspace{-2.5pt}I}(t;x_0 \Rightarrow 0)$ 
(d) Individual $f_I^{(n)}$ obtained from Eqs.(\ref{31_f12_s_Lap}) and (\ref{43_F1_s}).
The functions resulting in underestimation are represented as negative.}
\label{43_fig_F}
\end{figure}
%%%%%%%%%%%%%%%%%%%%%%%%%%%%%%%%%%%%%%%%%%%%%%%%%%%%%%%%%%%%%%%%%%%%%%%%%%%%%%%%%%%%%%%%%%%%%
As a next example, we consider the Ornstein-Uhlenbeck process, which is described by Eq.(\ref{21_Langevin}) with $U(x)=k (x-a)^2/2$.

Since the adjoint Fokker-Plank operator is given by
\begin{eqnarray}
\label{43_L}
  L^{+}_{FP}(y_0) = -\frac{y_0}{\tau} \frac{\partial}{\partial y_0} + D \frac{\partial^2}{\partial y_0^2},
\end{eqnarray}
where we introduce a new variable $y=x-a$,
we obtain
\begin{eqnarray}
\label{43_u}
  u^{\pm}_{s}(y_0) = e^{\frac{y_0^2}{4b^2}} \mathcal{D}_{-\tau s} \left( \pm \frac{y_0}{b} \right),
\end{eqnarray}
where $\tau=\gamma/k$, $b^2=D\tau$ and $\mathcal{D}_{v}(\cdot)$ is parabolic cylinder function \cite{OU_rep,OU_Sato}.
Requiring a regular behavior at $y_0 \to \pm \infty$, and using Eqs.(\ref{22_conv_Lap}) and (\ref{22_P_Lap}), 
we obtain
\begin{eqnarray}
\label{43_F1_s}
\tilde{F}_I(s;A &\Rightarrow B)  = 
\begin{cases}
  \frac{ u^{-}_{s}(A-a) }{ u^{-}_{s}(B-a) } & A \leq B \\
  \frac{ u^{+}_{s}(A-a) }{ u^{+}_{s}(B-a) } & A \geq B. \\ 
\end{cases}
\end{eqnarray}
Using Eqs.(\ref{31_f12_s_Lap}) and (\ref{43_F1_s}), we obtain the main results Eqs.(\ref{31_F2_s_2}) and (\ref{31_F2_t_inf}).

\begin{itemize}

  \item Comment to Eq.(\ref{31_F2_s_2})

    Unlike the cases of previous examples, we could not provide an analytical connection between our first solution    
    form and the result obtained by the EE method (see Appendix A). 
    Still, we expect they are closely related.
  \item Comment to Eq.(\ref{31_F2_t_inf})

    Unlike the cases of previous examples, the method of image generally does not work in the OU process, 
    because of the lack of reflection symmetry in the presence of harmonic potential.
    In this sense, our second solution form obtained through infinite filtration operations 
    may be seen as a generalization of the method of image.
  
\end{itemize}
In Fig.\ref{43_fig_F}, we show the performance of finite $N$ truncation approximation.

%%%%%%%%%%%%%%%%%%%%%%%%%%%%%%%%%%%%%%%%%%%%%%%%%%%%%%%%%%%%%%%%%%%%%%%%%%%%%%%%%%%%%%%%%%%%%
%%%%%%%%%%%%%%%%%%%%%%%%%%%%%%%%%%%%%%%%%%%%%%%%%%%%%%%%%%%%%%%%%%%%%%%%%%%%%%%%%%%%%%%%%%%%%

% 4-4
\subsection{Linearly Expanding Cage}

As a slightly different example, let us discuss the {\it Expanding Cage} problem, a first passage problem with moving absorbing boundaries \cite{Bray_EC,majumdar13,Red_EC}.
Here we assume boundaries to move linearly with time, for which we can make use the result in Sec.4.2.

We consider free Brownian motion with two absorbing boundaries, whose locations depend on time as follows:
\begin{eqnarray}
\label{44_boundary}
  L_{0}(t) = v_0 t, \, L_{L}(t) = L + v_L t,
\end{eqnarray}
where $L_{0}$ and $L_{L}$ denote positions of absorbing boundaries, with $v_{0}$ and $v_{L}$ being their velocities.
Subscript $0$ or $L$ in Eq.(\ref{44_boundary}) represents the initial position of each absorbing boundary at time $t=0$.
Considering the time dependence of boundary conditions,
let us rewrite FPT probability notation as $F_{I}(t_0,A(t_0) \Rightarrow t,B(t))$, 
where the right arrow with double lines indicates the first passage process for a random walker positioned at $x=A(t_0)$ at time $t_0$ to reach the absorbing boundary $x=B(t)$ at time $t$.
$F_I$ for this model can be expressed using that for the biased diffusion model (Sec.4.2.),
because we cannot distinguish a process for a biased random walker to approach (leave) a fixed boundary from a process, in which a linearly moving boundary approaches (leaves) an unbiased random walker \cite{guide}.
From Eq.(\ref{42_F1_t}), $F_{I}(t_0,A(t_0) \Rightarrow t,B(t))$ for this model can be expressed as 
\begin{equation}
\label{44_F1}
 F_I(t_0,A(t_0) \Rightarrow t,B(t)) = \frac{|B(t_0)-A(t_0)|}{\sqrt{4 \pi D (t-t_0)^3}} 
   \exp{\left[ -\frac{(B(t)-A(t_0))^2} {4D(t-t_0)} \right]}.
\end{equation}
Eq.(\ref{44_F1}) shows that $F_{I}(t_0,A(t_0) \Rightarrow t,B(t))$ cannot be generally expressed as a function of $t-t_0$ under the boundary conditions Eq.(\ref{44_boundary}).
Therefore, we cannot compute $f^{(n)}$ in the Laplace domain as in the case of other examples.
Thus, we rewrite the recursive formula of $f_I^{(n)}$ in the time domain
\begin{eqnarray}
  \label{44_f0}
f_I^{(0)}(0,x_0\Rightarrow t,B(t)) = F_{I}(0,x_0\Rightarrow t,B(t)) ,
\end{eqnarray}
\begin{eqnarray}
\label{44_fn}
  f_I^{(n)}(0,x_0 \Rightarrow t,B_1(t)) = 
   \int_{0}^{t} d\tau_n f_I^{(n-1)}(0,x_0 \Rightarrow \tau_n,B_2(t)) F_I(\tau_n,B_2(\tau_n) \Rightarrow t,B_1(t)) ,
\end{eqnarray}
where $B(t)=L_0(t)$ or $L_L(t)$ in Eq.(\ref{44_f0}) and $(B_1(t),B_2(t))=(L_0(t),L_L(t))$ or $(L_L(t),L_0(t))$ in Eq.(\ref{44_fn}).
The computation requires numerical integrations.
However, our method has no velocity restriction in Eq.(\ref{44_boundary}) and applies to any absorbing boundary condition with different velocities.
$F_{I\hspace{-2.5pt}I}$ can be obtained from Eqs.(\ref{31_F2_t_inf}), (\ref{44_f0}) and (\ref{44_fn}), where we assume the convergence of summation in Eq.(\ref{31_F2_t_inf}).

\begin{figure}[h]
  \begin{center}
  \includegraphics[width=150mm]{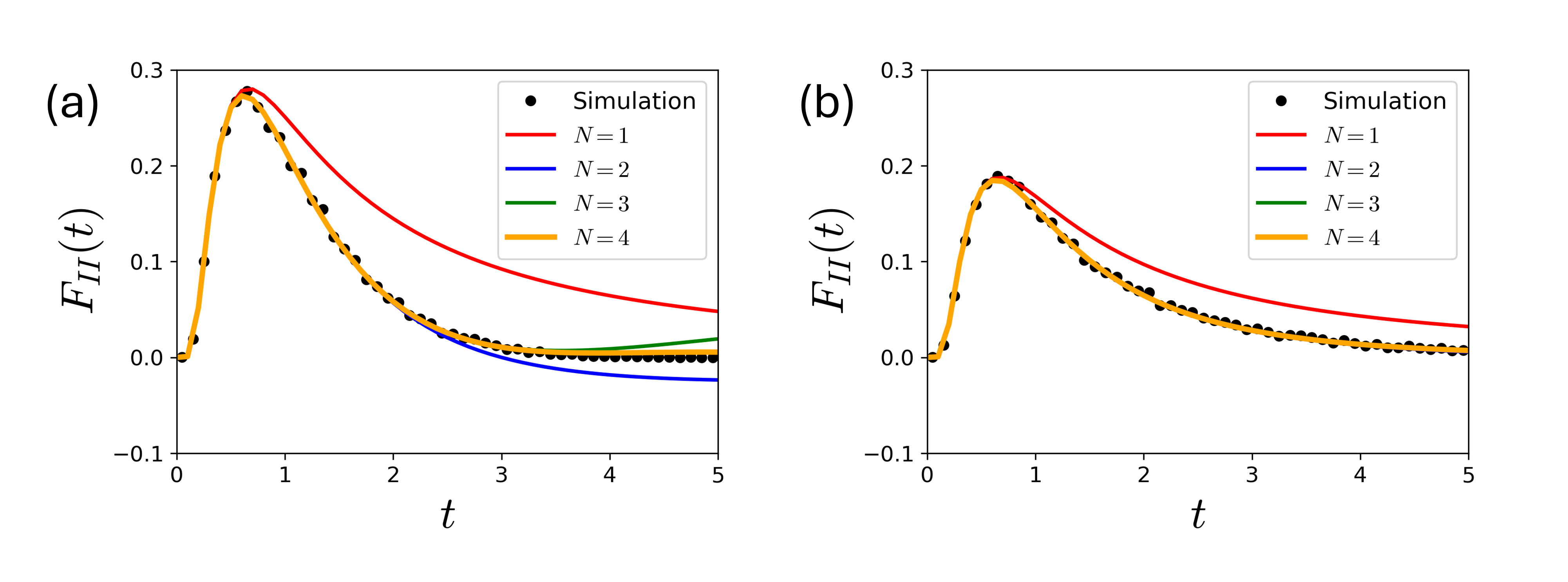}
  \caption{Comparison between the method of filtration (solid lines) and simulation results (dots).
  Results using the method of filtration are obtained from Eqs.(\ref{31_F2_t_inf}), (\ref{44_f0}) and (\ref{44_fn}).
  (a) Result for {\it Expanding Cage}: where $x_0=2$, $L=3$, $v_0=-0.2$, $v_L=0.1$.
  (b) Result for {\it Shrinking Cage}: where $x_0=2$, $L=3$, $v_0=0.2$, $v_L=-0.1$.
  Simulation results are based on $10^5$ independent realizations.
  }
  \label{44_Fig_F}
  \end{center}
\end{figure}

We performed numerical simulations for two cases: 
(1){\it Expanding Cage} ($v_0$ $<$ 0, $v_L$ $>$ 0), and
(2){\it Shrinking Cage} ($v_0$ $>$ 0, $v_L$ $<$ 0)
, and compared with our filtering theory.
As illustrated in Fig.\ref{44_Fig_F}, our method captures well the FPT distribution obtained in simulation with a small number $N$ of filtration.
%%%%%%%%%%%%%%%%%%%%%%%%%%%%%%%%%%%%%%%%%%%%%%%%%%%%%%%%%%%%%%%%%%%%%%%%%%%%%%%%%%%%%%%%%%%%%
%%%%%%%%%%%%%%%%%%%%%%%%%%%%%%%%%%%%%%%%%%%%%%%%%%%%%%%%%%%%%%%%%%%%%%%%%%%%%%%%%%%%%%%%%%%%%

%5
\section{Summary}

In this paper, we have proposed the method of filtration to obtain the FPT solution with two absorbing boundaries using only the solutions with one boundary.
We have demonstrated that the proposed method gives two solution forms, the EE method-like solution from finite filtration operations or the image method-like solution from infinite operations.
We have shown that, for free and biased diffusion, our first and second solution forms exactly match the EE method solutions and the method of image solutions, respectively.
However, the situation is different for the OU process.
As for our first form of solution, we could not so far prove its equivalence to that obtained by the EE method. 
Although we expect a close connection between the two, their exact relation in general stochastic processes remains to be ascertained.
As for our second form of solution, we emphasize that, for the OU process, the method of image generally does not work in the presence of harmonic potential because of the lack of reflection symmetry.
In this sense, our method with infinite filtration operations may be a sort of generalization for the method of image.

The FPT distribution calculated from the two solutions takes the form of series expansion.
However, these two expansions are markedly different.
To illustrate the point, we plot in Figs.\ref{41_fig_F},\ref{42_fig_F}, and \ref{43_fig_F}, the finite $N$ truncated approximations for both cases.
Since the first term in the EE method describes the slowest mode, truncation with a few modes provides a good description for the long-time regime, but not for the short-time regime. 
As we have already seen, such a trend is reversed for the image method, where the truncation at $N=2$ (or even smaller) is already sufficient to describe the short-time regime.
In this sense, the two solutions obtained from our method may be considered to be complementary to each other.

We expect that the proposed method of filtration finds application in a wide array of domains.
We have demonstrated its applicability to the {\it Expanding Cage} problem.
Another interesting application may be found in the non-Markovian process, where the challenge is to find an appropriate way to incorporate history-dependent memory into formalism \cite{FPT_Saka,Jeon,kardar07,kardar10,Benichou}.

\section{Acknowledgements}
We thank E.Carlon (KU Leuven) for the useful discussion.
This work was supported by the Aoyama Gakuin University Research Institute “Early Eagle” grant program for the promotion of research by early career researchers, and JSPS KAKENHI (Grant Nos.JP23H00369, JP23H04290, and JP24K00602).

\appendix
\newpage
% A
\section{Method of Eigenfunction Expansion}

For completeness, we briefly discuss a method to obtain FPT distribution with two absorbing boundaries $F_{I\hspace{-2.5pt}I}$ with the {\it Method of Eigenfunction Expansion}.
Here, we explain the case of the Ornstein-Uhlenbeck process \cite{OU_enrico, tpt_asy}. 
Other simpler cases, free diffusion and biased diffusion, can be analyzed with a similar procedure as well, see the concrete solution form Eqs.(27) and (34), respectively \cite{guide}.

For Brownian motion under harmonic potential $U(x)=k (x-a)^2/2$, 
the occupation probability $P(x,t)$ obeys the Fokker-Plank equation (\ref{21_FP}).
By making the separation of variables $P(x,t)=e^{-\frac{(x-a)^2}{4b^2}} Y(x)\Phi(t)$,
we obtain independent equation for position-dependent part $Y(x)$ and time-dependent part $\Phi(t)$
\begin{eqnarray}
\label{B_Yx}
  Y''(x) + \left( \frac{1}{2b^2} - \frac{(x-a)^2}{4b^4} + \lambda \right) Y(x) = 0,
\end{eqnarray}
\begin{eqnarray}
\label{B_Pt}
  \dot{\Phi}(t) = -\lambda D \Phi(t),
\end{eqnarray}
where prime and overdot represent spatial and temporal differentiation, and  $\lambda$ is a separation constant.
The solution of the time-dependent part can be easily found as
\begin{eqnarray}
\label{B_Pt_sol}
  \Phi(t) = e^{-s {t}/{\tau}},
\end{eqnarray}
where $s=b^2\lambda$.
Two independent solutions (even and odd) of Eq.(\ref{B_Yx}) are known \cite{HofM} as 
\begin{eqnarray}
\label{B_Yx_sol}
  Y(x) = 
\begin{cases}
  Y_{even}(x) = e^{-\frac{(x-a)^2}{4b^2}} 
  {}_1F_1 \left( -\frac{s}{2} ; \frac{1}{2} ; \frac{(x-a)^2}{2b^2}\right) \\
  Y_{odd}(x) = \frac{x-a}{b} e^{-\frac{(x-a)^2}{4b^2}} 
  {}_1F_1 \left( -\frac{s}{2} + \frac{1}{2} ; \frac{3}{2} ; \frac{(x-a)^2}{2b^2}\right),
\end{cases}  
\end{eqnarray}
where ${}_1F_1(a;b;z)$ is the Kummer confluent hypergeometric function.
Aside from the normalization, the general solution can be written as a linear combination of Eq.(\ref{B_Yx_sol}) as follows:
\begin{eqnarray}
\label{B_Yx_solG}
  Y(x) = Y_{even}(x) + A Y_{odd}(x),
\end{eqnarray}
where $A$ is determined by boundary conditions.

Imposing boundary conditions $Y(0)=Y(L)=0$, we find the constant $A=-{Y_{even}(L)}/{Y_{odd}(L)}$,
and the following equation
\begin{eqnarray}
\label{B_Yx_U}
  Y_{even}(0)}{Y_{odd}(L) - Y_{even}(L)}{Y_{odd}(0) = 0.
\end{eqnarray}
This equation yields a discrete spectrum $\{s_n\}=(s_1,s_2, \dots ,)$ allowed by the boundary conditions, where we assume the order $(s_1<s_2<\cdots)$.

The solution of Eq.(\ref{21_FP}) can be expressed as a linear combination of eigenfunctions
\begin{eqnarray}
\label{B_P_LC}
  P(x,t)=e^{-\frac{(x-a)^2}{4b^2}} \sum_n c_n Y_n(x)\Phi_n(t),
\end{eqnarray}
where the constants $c_n$ are determined from initial condition $P(x,0)=\delta(x-x_0)$ using the orthogonality of $Y_n(x)$ and normalizing $Y_n(x_0)$ as $\int_{0}^{L} dx Y_n^2(x_0) = 1$
\begin{eqnarray}
\label{B_P_c}
  c_n=e^{-\frac{(x_0-a)^2}{4b^2}} Y_n(x_0).
\end{eqnarray}
We finally obtain the occupation probability under absorbing boundary conditions
\begin{eqnarray}
\label{B_P_final}
  P(x,t)=e^{-\frac{(x-a)^2-(x_0-a)^2}{4b^2}} \sum_n Y_n(x_0) Y_n(x) e^{-s_n t/\tau}.
\end{eqnarray}

FPT probability is equal to the flux of $P(x,t)$ to the absorbing boundary.
The flux $j$ of Fokker-Plank equation Eq.(\ref{21_FP}) is expressed as 
\begin{eqnarray}
\label{B_Flux}
   j(x,t) = 
     -\left[ \frac{1}{\gamma} \frac{\partial U(x)}{\partial x} + D\frac{\partial}{\partial x}\right] P(x,t).
\end{eqnarray}
Therefore, we obtain
\begin{eqnarray}
\label{B_Fpt}
  F_{I\hspace{-2.5pt}I} (t;x_0 \Rightarrow 0) = j(0,t) = 
   \left. 
      -\left[ \frac{1}{\gamma} \frac{\partial U(x)}{\partial x} + D\frac{\partial}{\partial x}\right] P(x,t)
   \right|_{x=0}.
\end{eqnarray}
%

%B
\section{Inverse Laplace transform of Eq.(\ref{41_F2_s_N})}
In Sec4.1, we have shown that the method of filtration gives the same solution as that obtained from the EE method by performing the inverse Laplace transform of the first solution form Eqs.(\ref{31_F2_s_N}) or (\ref{31_F2_s_2}).
Here, let us discuss the way to solve Eqs.(\ref{31_F2_s_N}) or (\ref{31_F2_s_2}), which is to perform inverse Laplace transform, specifically in the free diffusion case.
The biased diffusion case can also be calculated using a similar procedure.

As a preparation to solve Eq.(\ref{41_F2_s_N}), let us consider the inverse Laplace transform for 
$\tilde{f}(s) = \frac{\sinh{(B\sqrt{s})}}{\sinh{(A\sqrt{s})}} = \frac{\sin(iB\sqrt{s})}{\sin(iA\sqrt{s})}$.
It can be computed by residue theorem as 
\begin{eqnarray}
  f(t) & = \mathcal{L}^{-1} \left[ \tilde{f}(s) \right]  \\ \nonumber
    & = \sum_{n=0}^{\infty} Res_{(s=s_n)} \left[ \frac{\sin(iB\sqrt{s})}{\sin(iA\sqrt{s})} e^{ts} \right],
\label{B_iLt}
\end{eqnarray}
where $Res_{(s=s_n)}[\tilde{f}(s)e^{ts}]$ represents for residue of $\tilde{f}(s)e^{ts}$, and $\tilde{f}(s)e^{ts}$ has simple poles when $s_n=-(\frac{n\pi}{A})^2$.
Each term in the summation is
\begin{eqnarray}
\label{B_Res}
  Res_{(s=s_n)} \left[ \frac{\sin(iB\sqrt{s})}{\sin(iA\sqrt{s})} e^{ts} \right]
   &= \lim_{s \to s_n} \frac{s-s_n}{\sin(iA\sqrt{s})} \sin(iB\sqrt{s}) e^{ts} \\
   &= 2 (-1)^{n+1} \frac{n\pi}{A^2} \sin \left(\frac{B}{A} n\pi \right) e^{s_nt},
\end{eqnarray}
where we use L'Hospital's rule.
Thus,
\begin{eqnarray}
\label{B_f_t}
  f(t) &= \sum_{n=0}^{\infty} 2 (-1)^{n+1} \frac{n\pi}{A^2} \sin \left(\frac{B}{A} n\pi \right) e^{s_nt}.
\end{eqnarray}
Using Eqs.(\ref{41_F2_s_N}) and (\ref{B_f_t}), we finally obtain Eqs.(\ref{41_F2_t_N}) and (\ref{41_F2_t_2}).
We emphasize that the origin of summation over $m$ in Eq.(\ref{41_F2_t_2}) (also in Eq.(\ref{41_F2_t_N})) is the contribution of the infinite number of poles in Eq.(\ref{41_F2_s_N}), not the repeated filtration operations.

\end{document}